\documentclass[11pt]{paper}

\usepackage{mathrsfs}
\usepackage{amsmath,amssymb,amsthm}
\usepackage{cite}
\usepackage{color}
\usepackage{hyperref}
\usepackage{tikz,pgf}
\usepackage{cancel}
\usepackage[mathscr]{euscript}  
\usepackage{mathtools} 
\numberwithin{equation}{section} 
\usepackage{todonotes}
\usepackage{amsmath,amssymb}
\usepackage{hyperref}
\usepackage{stackengine,scalerel}

\newcommand\DDDag{%
  \sbox0{\ddag}\stretchrel*{%
  \stackengine{-.6\ht0}{\ddag}{\ddag}{O}{c}{F}{F}{S}}{\ddag}%
}

\numberwithin{equation}{section}

\newcommand{\bse}{\begin{subequations}}
\newcommand{\ese}{\end{subequations}}

\def\undertilde#1{\mathord{\vtop{\ialign{##\crcr
$\hfil\displaystyle{#1}\hfil$\crcr\noalign{\kern1.5pt\nointerlineskip}
$\hfil\widetilde{}\hfil$\crcr\noalign{\kern-6.5pt}}}}}
\def\underhat#1{\mathord{\vtop{\ialign{##\crcr
$\hfil\displaystyle{#1}\hfil$\crcr\noalign{\kern1.5pt\nointerlineskip}
$\hfil\widehat{}\hfil$\crcr\noalign{\kern-6.5pt}}}}}
\def\underbar#1{\mathord{\vtop{\ialign{##\crcr
$\hfil\displaystyle{#1}\hfil$\crcr\noalign{\kern1.5pt\nointerlineskip}
$\hfil\bar{}\hfil$\crcr\noalign{\kern-6.5pt}}}}}

\newcommand{\be}{\begin{equation}}
\newcommand{\ee}{\end{equation}}

\newcommand{\bea}{\begin{eqnarray}}
\newcommand{\eea}{\end{eqnarray}}

\title{
The 
 Darboux-KP system as an integrable Chern-Simons multiform theory 
 in infinite dimensional space
\\
}

\author{Jo\~{a}o Faria Martins$^\dagger$, Frank W Nijhoff$^\ddagger$ and Daniel Riccombeni$^{\DDDag}$ \\
 {\small  School of Mathematics, University of Leeds, Leeds LS2 9JT, UK }\\
\small $\dagger$ \texttt{J.FariaMartins@leeds.ac.uk} \\ 
\small $\ddagger$ \texttt{F.W.Nijhoff@leeds.ac.uk}\\
\small $\DDDag$ \texttt{D.A.Riccombeni@leeds.ac.uk}\\
}

\newcommand{\ignore}[1]{}
\begin{document}

\maketitle

\begin{abstract}
In a previous paper by one of the authors, a Lagrangian 3-form structure was 
established for a generalised Darboux system, originally describing orthogonal curvilinear coordinate systems, which encodes the 
Kadomtsev-Petviashvili (KP) hierarchy. Here a hierarchy of Lagrangian multiforms 
is established for the same system, viewed as a hierarchy of Chern-Simons 
actions in an infinite-dimensional space of Miwa variables, constituting 
the variational form of a universal 3D integrable system embedded in this infinite-dimensional
space.

\paragraph{Keywords:} Integrable system, multi-dimensional consistency, Lagrangian 
multiforms, KP hierarchy, Darboux system, Chern-Simons action.  
\end{abstract}

\section{Introduction}\label{S:Intro}
 The notion of Lagrangian multiforms was introduced in \cite{LobbNij2009} as a way 
 to incorporate the key integrability aspect of multidimensional 
 consistency (MDC), meaning the coexistence of multiple compatible equations on one and the same 
 dependent variable, into a variational framework. In that framework, the 
 Lagrangians are differential or difference $d$-forms of nontrivial codimension, 
 and the action is a functional of not only the variational fields, but also of 
 the surfaces over which the $d$-form is integrated in a `multi-time' space of 
 independent variables of arbitrary dimension. For this to work, the Lagrangian 
 components of the $d$-form ($d$ being related to the dimensionality of the 
 equations in the system) must be very special, in fact `integrable', and they emerge 
 as solutions of a generalised Euler-Lagrange (EL)  system of equations, cf. e.g. \cite{SurVerm,Sleigh2020}. It was shown 
 in subsequent work that many known integrable systems (both continuous and discrete) 
 admit a Lagrangian 
 multiform structure, (cf. e.g. \cite{HJN2016} and references therein).

In \cite{Nij2023} a Lagrangian 3-form action was presented for a generalised 
Darboux system (a system originally describing conjugate nets of orthogonal curvilinear 
coordinates) embedded in a infinite-dimensional space of `Miwa variables', \cite{Miwa}, 
which encodes the entire Kadomtsev-Petviashvili (KP) hierarchy, \cite{MartinezKonop}. 
While a Lagrangian 3-form structure was given in \cite{Sleigh2022} for the KP hierarchy in terms of the usual 
description in terms of pseudo-differential operators, the 'Darboux-KP' system represents 
the KP hierarchy in terms of some compounded parameter-dependent 
variables, called Miwa Variables (cf. also \cite{Miwa,Nij1987}). 
The 3-form action takes the form 
\begin{equation}\label{eq:action}
{\sf S}[\mathbf{B}(\boldsymbol{\xi});\mathcal{V}]=\int_{\mathcal{V}} {\sf L}
\end{equation}
which should be considered as a functional not only of the field variables 
$\mathbf{B}$, but also of the 3-dimensional hypersurfaces $\mathcal{V}$ 
embedded in an infinite-dimensional space of independent variables 
$\boldsymbol{\xi}=\{\xi_{p_i}\}_{i\in I}$ (where $I$ is some well-chosen 
index set of cardinality at least 3, but later on we will we take $I=\mathbb{Z}$). In the present context the variables $\xi_{p}$ are 
labelled by a set of continuous parameters $\{ \boldsymbol{p}=(p_j)_{j\in I} \}$, 
which themselves can be complex valued, and they correspond to the so-called 
Miwa variables in the KP theory, cf. \cite{Miwa}. 
The `Lagrangian' \eqref{eq:action} $\mathsf{L}$ is here a differential 3-form 
\begin{align} \label{eq:gen3form} 
\mathsf{L}= \sum_{i,j,k\in I} \mathcal{L}_{p_i,p_j,p_k}\,{\rm d}\xi_{p_i}\wedge {\rm d}\xi_{p_j} 
\wedge{\rm d}\xi_{p_k}\ , 
\end{align} 
with Lagrangian components $\mathcal{L}_{p_i,p_j,p_k}$ given by 
\begin{align}\label{eq:Lagr}
\mathcal{L}_{pqr} = & \tfrac{1}{2}\left(B_{rq}\partial_{\xi_p}B_{qr}-B_{qr}\partial_{\xi_p}B_{rq} \right)  
  +\tfrac{1}{2}\left(B_{qp}\partial_{\xi_r}B_{pq}-B_{pq}\partial_{\xi_r}B_{qp} \right) \nonumber \\ 
   & +\tfrac{1}{2}\left(B_{pr}\partial_{\xi_q}B_{rp}-B_{rp}\partial_{\xi_q}B_{pr} \right)  
   + B_{rp} B_{pq} B_{qr} - B_{rq}B_{qp} B_{pr}.
\end{align}

The outstanding feature, proven in \cite{Nij2023}, is that the differential of the Lagrangian 3-form 
\[ {\rm d}\mathsf{L}= \sum_{i,j,k,l\in I} \mathcal{A}_{p_i,p_j,p_k,p_l}\,
{\rm d}\xi_{p_i}\wedge {\rm d}\xi_{p_j}\wedge{\rm d}\xi_{p_k}\wedge{\rm d}\xi_{p_l}\ , \] 
has a `double zero' (to be explained below) on solutions of a multidimensionally consistent system of 
equations 
\begin{equation}\label{eq:Beqs}  
\frac{\partial B_{p_i,p_j}}{\partial \xi_{p_k} }= B_{p_i,p_k}B_{p_k,p_j}\ , \quad \textrm{ whenever 
$p_j$,  $p_j$,  $p_k$ are different}\  , 
\end{equation}
which is a generalisation of the Darboux systems describing conjugate nets for 
curvilinear coordinates, \cite{Darboux}. As a consequence, the set of generalised Euler-Lagrange (EL) 
equations arise from the criticality condition $\delta{\mathrm d}{\mathsf L}=0$ (in the language of the 
corresponding variational bicomplex,\cite{Dickey}), and the latter condition produces the whole multidimensionally 
consistent set of Darboux equations from a single variational principle. 
Finally, as shown in \cite{Nij2023}, cf. also \cite{MartinezKonop}, the set of 
Darboux equations in terms of Miwa variables is equivalent to the entire 
KP hierarchy of equations, while the integrability of the original finite-dimensional system of Darboux and Lam\'e, 
\cite{Lame}, 
was studied in \cite{Zakharov96}. 

In fact, the fields $B_{pq}$ can be expressed in terms of the KP $\tau$-function, 
$$ \tau(t_1,t_2,\cdots)=\tau(\{\xi_{p_i}\}_{i\in I}; \{ n_{p_i}\}_{i\in I}), $$ 
in an extended 
space with discrete variables $n_{p_i}$ associated with the parameters $p_i$, $i\in I$, 
as 
$$ B_{p_ip_j}=\frac{X_{p_ip_j}\tau}{(p_i-p_j)\tau} .$$
Here $X_{pq}$ is the $\frak{gl}(\infty)$ vertex operator, \cite{DKJM1983}, which can be expressed 
as 
$$ X_{p_ip_j}=\left[\frac{\prod_{l\in I\atop l\neq j} (p_l-p_j)^{n_{p_l}}}
{\prod_{l\in I\atop l\neq i} (p_l-p_i)^{n_{p_l}}}\right]\, T_{p_j}T_{p_i}^{-1} , $$ 
in terms of discrete variables, where $T_{p_i}$ and $T_{p_j}$ are the elementary lattice shift operators in the variables $n_{p_i}$ and 
$n_{p_j}$ respectively. Alternatively, we have 
$$
   X_{p_ip_j} =\exp \left(\sum_{l\in I\atop l\neq i} \frac{\xi_{p_l}}{p_i-p_l} 
   +\sum_{l\in I\atop l\neq j} \frac{\xi_{p_l}}{p_l-p_j}\right)  
   \exp\left( \sum_{l=1}^\infty \frac{1}{l}\left(\frac{1}{p_i^l}-\frac{1}{p_j^l} \right)
   \frac{\partial}{\partial t_l} \right)
$$
in terms of the continuous variables $t_j$ of the KP hierarchy, in terms of which the $\tau$-function 
obeys the set of Hirota bilinear equations, cf. \cite{Sato,MJD}. Here the Miwa variables, $\xi_p$, are related to the higher time variables, $t_j$,  in the conventional way of writing the KP hierarchy by 
$$ \frac{\partial\tau}{\partial \xi_p}=\sum_{j=1}^\infty \frac{1}{p^{j+1}}\frac{\partial\tau}{\partial t_j}\ , \quad 
T_{p_i}\tau=\tau\left(  \big\{ t_l-\frac{1}{lp_i^l}\big\}_{l \in \mathbb{Z}^+}\right).
$$ 
Furthermore, there is the relation 
$$  \frac{\partial\tau}{\partial \xi_p}= \left(T_p^{-1}\frac{d}{dp} T_p \right)\tau $$
connecting the two realisations.

\section{Higher-dimensional Chern-Simons theories} 

The aim of this paper is to show that the Lagrangian \eqref{eq:Lagr} for the Darboux-KP system 
arises from an infinite-dimensional matrix-valued Chern-Simons action \cite{CS71,CS}. This would seem to place the notion of integrability in the broad
domain of topological 
field theories, but as we shall see this is only in appearance.
A connection between topological field 
theory, and in particular the Chern-Simons action, was demonstrated in \cite{Martina1997}, cf. also 
\cite{Martina2001}. In these papers it was shown that the CS action, by specific choices of  
(finite-dimensional) gauge groups lead to cases of integrable models (notably of Davey-Stewartson 
and Ishimori type) in 2+1 dimensions, but this connection does not seem to fully reveal the inherent 
integrability of those models. 

A decade earlier, a connection
between 1+1-dimensional integrable hierarchies and topological field theories 
(of the Wess-Zumino-Witten type) was pointed out in \cite{Nij1987}. Here the 
coupling constant in front of the topological term in the action contains essentially 
the spectral parameter. Similar coupling constants were rediscovered in the 
work by Costello, Witten and Yamazaki, cf. \cite{Costello,CostelloYam}, in the context of what is called 4D Chern-Simons theory 
describing integrable field theories in 1+1 dimensions (with two of the real 
dimensions being reserved for the complex spectral parameter).
It would be interesting to determine how these 
various connections between integrability and topological field theory are interrelated, and moreover how the action 
incorporates the specifics of certain solution classes\footnote{Noting that in integrable systems the 
behaviour of specific solutions is linked to the structure of the manifolds in spectral space, and this 
should be linked to the manifolds over which the 4D CS Lagrangian is integrated. Thus the question arises of 
how the base manifold in the latter theory reflects the nature of the solutions, and whether the corresponding action is genuinely universal (i.e., solution-independent).}.

The conventional Chern-Simons theory over a Lie algebra $\frak{g}$, with 
associated gauge group $G$, involves a $\frak{g}$-valued gauge connection 
$\boldsymbol{A}$, which is a differential 1-form, and the associated curvature 2-form,  
$\boldsymbol{F}={\rm d}\boldsymbol{A}+\boldsymbol{A}\wedge\boldsymbol{A}$. 
Here, however, we consider matrix-valued gauge fields only, where the gauge groups of interest are the 
general linear groups, $GL(n,\mathbb{R})$, in whose Lie algebra, $\mathfrak{gl}(n,\mathbb{R})$, we consider 
the matrix trace ${\rm Tr}$, and where the wedge product $\boldsymbol{A}\wedge\boldsymbol{A}$  is 
evaluated via the matrix product, and not via the Lie bracket.

The standard CS Lagrangians in dimensions 3 and 5 read 
\begin{subequations}\begin{align}
CS_3& = {\rm Tr}\left(\boldsymbol{A}\wedge{\rm d}\boldsymbol{A}+\tfrac{2}{3} 
\boldsymbol{A} \wedge\boldsymbol{A}\wedge\boldsymbol{A}\right) \ , \label{de:CS3}\\ 
CS_5&= {\rm Tr}\left(\boldsymbol{A}\wedge{\rm d}\boldsymbol{A}\wedge{\rm d}\boldsymbol{A}
+\tfrac{3}{2} 
\boldsymbol{A} \wedge\boldsymbol{A}\wedge\boldsymbol{A}\wedge{\rm d}\boldsymbol{A}+\tfrac{3}{5}\boldsymbol{A} \wedge\boldsymbol{A}\wedge\boldsymbol{A}\wedge\boldsymbol{A}\wedge\boldsymbol{A}\right) \ ,\label{de:CS5} 
\end{align}\end{subequations}
and they are defined through the property that 

\begin{subequations}\label{eq:CSs}\begin{align}
    {\rm d}CS_3& ={\rm Tr}\left(\boldsymbol{F}\wedge\boldsymbol{F}\right)\  , \label{eq:CS3} \\  
    {\rm d}CS_5& ={\rm Tr}\left(\boldsymbol{F}\wedge\boldsymbol{F}\wedge\boldsymbol{F}\right)\  . \label{eq:CS5} 
\end{align}\end{subequations}
In hindsight, coming from the variational principle for the multiform theory, the 
relations \eqref{eq:CSs} are nothing but the double, respectively 
triple zero conditions for the Euler-Lagrange equations $\boldsymbol{F}=0$. 
The general higher form of the CS Lagrangians are given by the formula which is a particular case of  \cite[Lemma 3.3]{CS71}, \begin{equation}\label{eq:hi-CS} 
CS_{2n+1}= (n+1)\int_0^1 {\rm d}\lambda\, {\rm Tr}\left( \boldsymbol{A}\wedge 
\boldsymbol{F}_\lambda^{\wedge n}\right)\ , \quad {\rm where}\quad 
\boldsymbol{F}_\lambda:= \lambda{\rm d}\boldsymbol{A}+\lambda^2\boldsymbol{A}\wedge\boldsymbol{A}\ , 
\end{equation}
see also \cite{Zanelli}. They obey\footnote{Note that in dimension 1, we have $CS_1={\rm Tr}(\boldsymbol{A})$, with ${\rm d}CS_1={\rm Tr}(\boldsymbol{F})$. }
\begin{equation}{\rm d}CS_{2n+1}={\rm Tr}\left( \boldsymbol{F}^{\wedge (n+1)}\right)\ ,  \end{equation}
where the latter expressions are $2n+2$-forms, whose variational derivative (using the 
multiform EL equations in the 
language of the variational bi-complex), are given by  \[ \delta {\rm d}CS_{2n+1}=(n+1){\rm Tr}\left(\boldsymbol{F}^{\wedge (n)}\wedge 
\delta\boldsymbol{F}\right)\ , \]
which vanishes whenever $\boldsymbol{F}=0$. The latter would correspond to the usual variational 
equations in conventional CS theory, cf. e.g. \cite{Wernli}, but as we will see later, in our setting $\boldsymbol{F}=0$ is too stringent a condition. In fact, 
as we will work later with a restricted set of fields the corresponding equations of motion
are slightly weaker than the standard zero-curvature condition.

In all these CS theories, we fix the dimensionality of the (2$n$+1)-dimensional manifold $\mathcal{M}_{2n+1}$ over 
which the Lagrangians are integrated through 
\begin{equation}\label{eq:CSaction} 
\mathcal{A}^{CS}_{2n+1}=\int_{\mathcal{M}_{2n+1}} CS_{2n+1}\ .
\end{equation} 

So far we have summarised conventional CS theory, which is by itself not an integrable 
theory. In fact, the various CS actions \eqref{eq:hi-CS} are separate theories living 
each in a different dimension. What we aim for is to create a \textit{universal} and `integrable' CS theory which lives in an a priori infinite-dimensional space, and 
where all CS actions are co-existing and on the same footing. For this we need 
gauge connections which are compatible in all these dimensions, and hence possess common 
solutions. This will lead necessarily to a MDC system, and the 
corresponding CS theory will have a Lagrangian multiform structure. 

\section{Higher Lagrangian Multiforms from CS Lagrangians}

In order to make a connection between the CS theory and Lagrangian multiforms 
we need to specify the gauge field $\boldsymbol{A}$ in such a way that the 
Lagrangian 3-form \eqref{eq:gen3form} with \eqref{eq:Lagr} is recovered. 
Although it is not obvious that such an identification exists, it turns out that 
the following choice achieves this\footnote{An alternative choice would be 
to consider the 1-form, below instead of \eqref{eq:B},
$$ \boldsymbol{C}= \sum_{k,l \in \mathbb{Z}} -C_{kl}\, {\rm d} \xi_l \, E_{k,l}$$ 
 which leads to the same system of equations 
for the coefficients $C_{ij}$ as for the fields $B_{ij}$. }
\begin{equation}
  \label{eq:B} 
\boldsymbol{B}=\sum_{k,l \in \mathbb{Z}} B_{kl}\, {\rm d} \xi_k \, E_{k,l}\ .
\end{equation} 
Here $B_{kl}=B_{kl}\left( (\xi_k)_{k \in \mathbb{Z}} \right )$, and we have simplified the notation by replacing $\xi_{p_i}$ simply by 
$\xi_i$, and $B_{p_k,p_l}$ simply by $B_{k,l}$, (we will come back to the role of the parameters $p_i$ later)\footnote{Note that, for convenience, we took label set $I=\mathbb{Z}$, however in principle $I$ could even be a continuous set of labels.}. 
Each $E_{kl}$ is an element of the space of all matrices $(M_{ab})_{a,b \in \mathbb{Z}}$, indexed by integers, with $(E_{kl})_{ab}=\delta_{k,a} \delta_{l,b}$.  The sum in \eqref{eq:B} is infinite, however it can be understood in the `completed graded sense', for the ${\rm d} \xi_k \, E_{k,l}$ are linearly independent, and hence \eqref{eq:B} never leads to infinite sums of real numbers. 
Note that $E_{k,l}E_{m,n}=\delta_{l,m}E_{k,n}$ and  ${\rm Tr}(E_{k,l})=\delta_{k,l}$.  

Computing the curvature $\boldsymbol{F}$ associated with this gauge field 
$\boldsymbol{B}$, we get 
\begin{equation}
\boldsymbol{F}_B=\sum_{j, k,l \in \mathbb{Z}}  \big (\partial_{\xi_j} B_{kl} - B_{kj} B_{jl}\big) \, {\rm d} \xi_j \wedge {\rm d} \xi_{k} \, E_{k,l}\ ,
\end{equation}
where the coefficients  $\partial_{\xi_j} B_{kl} - B_{kj} B_{jl} $ are 
exactly of the Darboux form of \eqref{eq:Beqs}. 
With the choice \eqref{eq:B} of gauge field we can now compute the CS action, leading to the Lagrangian 3-form \eqref{eq:gen3form} together with \eqref{eq:Lagr}. The Lagrangian 3-form of the Darboux-KP system arises from the $CS_3(\boldsymbol{B})$ by direct computation, namely 
\[ \mathsf{L}^{(3)}= CS_3(\boldsymbol{B})= \sum_{i,j,k \in \mathbb{Z}} \mathcal{L}^{(3)}_{ijk} 
\,{\rm d} \xi_i \wedge {\rm d} \xi_j \wedge {\rm d} \xi_k\ ,  \] 
in which the coefficients $\mathcal{L}^{(3)}_{ijk}=\tfrac{2}{3!}\mathcal{L}_{p_ip_jp_k}$ of \eqref{eq:Lagr} including a prefactor for convenience.

Calculating the form of the right-hand side in \eqref{eq:CS3} we find 
\begin{multline}
{\rm Tr}(\boldsymbol{F}_B\wedge \boldsymbol{F}_B)=\\ 
\sum_{\substack{ j,k,l,m \in \mathbb{Z}\\
\text{all indices different}}}  \!\!\!\!\!\!\big (\partial_{\xi_j} B_{kl} - B_{kj} B_{jl}\big)  \big (\partial_{\xi_{m}} B_{lk} - B_{lm} B_{mk}\big) \, {\rm d} \xi_{j} \wedge {\rm d} \xi_{k} \wedge\, {\rm d} \xi_{m} \wedge {\rm d} \xi_{l}.
\end{multline}
In particular, this implies that ${\rm Tr}(\boldsymbol{F}_B\wedge \boldsymbol{F}_B)$,  has a double zero on the solutions of the generalised 
Darboux system in  \eqref{eq:Beqs}, which implies that the latter arises as the EL equations of the multiform action 
\eqref{eq:action} with \eqref{eq:gen3form}, as was asserted in \cite{Nij2023}. 

\paragraph{Remark:} Note that while ${\rm Tr}(\boldsymbol{F}_B \wedge \boldsymbol{F}_B)$ indeed has a double zero on the solutions of \eqref{eq:Beqs}, the form $\boldsymbol{F}_B\wedge \boldsymbol{F}_B$ does not necessarily have such a double zero when \eqref{eq:Beqs} holds, as this would require that the Darboux system also extends to the 
case that all three labelled variables are no longer distinct. 

\medskip

Inspired by this new point of view on the Chern-Simons origin of the Darboux-KP 
multiform action, it is natural to consider the higher order CS actions as well. Noting that the $CS_3$ provides a system of first order in the derivatives, 
the higher Chern-Simons actions will yield higher order Lagrangian multiforms, 
which haven't been considered yet  in the literature. 
These higher multiforms for the Darboux-KP system are directly obtained from the higher CS 
actions by using \eqref{eq:B}. Thus, we obtain the Lagrangian 5-form 
\begin{align} 
\mathsf{L}^{(5)} =& {\rm Tr}\left(\boldsymbol{B}\wedge{\rm d}\boldsymbol{B}\wedge{\rm d}\boldsymbol{B}
+\tfrac{3}{2} 
\boldsymbol{B} \wedge\boldsymbol{B}\wedge\boldsymbol{B}\wedge{\rm d}\boldsymbol{B}+\tfrac{3}{5}\boldsymbol{B} \wedge\boldsymbol{B}\wedge\boldsymbol{B}\wedge\boldsymbol{B}\wedge\boldsymbol{B}\right) \ ,\nonumber 
\\ 
& =\sum_{j,k,l,m,n\in \mathbb{Z}} \mathcal{L}^{(5)}_{jklmn} {\rm d} \xi_j \wedge {\rm d} \xi_{k}\wedge {\rm d} \xi_l \wedge {\rm d} \xi_{m}\wedge {\rm d} \xi_n \ , \label{eq:L5} 
\end{align} 
with
\begin{align}\label{eq:hi-Lagr}
\mathcal{L}^{(5)}_{jklmn} = & 
\tfrac{1}{5!}\sum_{j',k',l',m',n'\in \{j,k,l,m,n\}} \varepsilon_{j'k'l'm'n'}\Big[ 
B_{p_{j'},p_{l'}}(\partial_{\xi_{p_{k'}}}B_{p_{l'},p_{n'}})
(\partial_{\xi_{p_{m'}}}B_{p_{n'},p_{j'}}) \nonumber \\ 
& + \tfrac{3}{2}  B_{p_{j'},p_{k'}}B_{p_{k'},p_{l'}}B_{p_{l'},p_{n'}}(\partial_{\xi_{p_{m'}}}B_{p_{n'},p_{j'}})    \nonumber \\ 
&  +\tfrac{3}{5}  B_{p_{j'},p_{k'} B_{p_{k'},p_{l'}} B_{p_{l'}},p_{m'}} B_{p_{m'},p_{n'}}B_{p_{n'},p_{j'}} \Big]\  , 
\end{align} 
where $\varepsilon_{jklmn}$ is the 5-dimensional Levi-Civita symbol. 
As a consequence of the construction, the Lagrangian 5-form has the property that 
$${\rm d} \mathsf{L}^{(5)}={\rm Tr}\left(\boldsymbol{F}_B \wedge \boldsymbol{F}_B \wedge \boldsymbol{F}_B\right)\ . $$
This again leads to the fact that ${\rm d} \mathsf{L}^{(5)}$ has a triple zero on the solutions of the same Darboux system \eqref{eq:Beqs}, viewed as a multidimensionally consistent system in higher dimensions. (Note again that $\boldsymbol{F}_B \wedge \boldsymbol{F}_B \wedge \boldsymbol{F}_B$ does not necessarily have a triple zero on the solutions of \eqref{eq:Beqs}.)

Similarly, all higher Lagrangian multiforms $\mathsf{L}^{(2n+1)}$ of odd degree can be constructed in the same way, using the formula \eqref{eq:hi-CS}. In fact, the differential  
\begin{multline}
{\rm d}\mathsf{L}^{(2n+1)}={\rm Tr}\left ( \boldsymbol{F}_B^{\wedge n}\right )=\sum_{j_1,l_1, j_2,l_2,\dots,j_n,l_n \in \mathbb{Z}}\\ (\partial_{\xi_{j_1}} B_{l_n l_1} -B_{l_n j_1} B_{j_1 l_1})\, (\partial_{\xi_{j_2}} B_{l_1 l_2} -B_{l_1 j_2} B_{j_2 l_2}) \, \dots\,  (\partial_{\xi_{j_n}} B_{l_{(n-1)} l_n} -B_{l_{(n-1)} j_n} B_{j_n l_l}) \\ {\rm d}  \xi_{j_1} \, \wedge\, {\rm d}  \xi_{l_n}\, \wedge \,  {\rm d}  \xi_{j_2} \, \wedge  {\rm d}  \xi_{l_1}\, \wedge \, \dots \, \wedge\, {\rm d}\xi_{j_n} \, \wedge \, {\rm d} \xi_{l_{(n-1)} }, 
\end{multline}
has an $n$-fold zero on the solutions of the generalised Darboux-KP system by virtue of the fact that none of 
the indices in the coefficients of the elementary forms on the r.h.s. coincide.

At this point it is important to note that due to the MDC property of the Darboux-KP system 
all these higher order CS actions are compatible amongst each other in the sense that the 
resulting multiform EL equations arising from each degree $n$ are consistent. This allows us to consider all these 
higher order CS theories together and create a `universal' CS action in infinite-dimensional space generating 
the whole series.

\section{Discussion: the generating Chern-Simons multiform} 

We have established a hierarchy of Lagrangian multiforms associated with the 
generalised Darboux system \eqref{eq:Beqs}, arising from a particular system of higher 
CS Lagrangians. These higher Lagrangian multiforms, of arbitrary odd degree, are all
compatible in the sense that the relevant EL equations all produce the 
multidimensionally consistent Darboux system \eqref{eq:Beqs}. 
Although these CS Lagrangians generate these Lagrangian multiforms via the choice \eqref{eq:B} of 
the corresponding gauge fields, in contrast to the usual CS theory \cite[\S 1.3]{Wernli} a critical point of the action 
can arise even when $\boldsymbol{F}_B\neq 0$. This is because we are not considering gauge fields 
in `general position', and variations $\delta\boldsymbol{B}=t\boldsymbol{\eta}$ are consequently 
restricted, i.e.  $\boldsymbol{\eta}=\sum_{p,q} \eta_{p,q} {\rm d} \xi_p E_{pq}$. Thus, for the $n=1$ 
CS action, assuming that integration manifold $\mathcal{V}$ is closed, we obtain by applying Stokes theorem 
\begin{align*}
\tfrac{d}{dt}    \int_{\mathcal{V}} CS_3(\boldsymbol{B}+t\boldsymbol{\eta})_{t=0}=2\int_{\mathcal{V}} \sum_{a,b,c} (\partial_{\xi_a} B_{bc}-B_{ba}B_{ac}) \eta_{cb} \, {\rm d}\xi_a \wedge {\rm d}\xi_b \wedge {\rm d} \xi_c\ . 
\end{align*}
Hence, a critical point of the action exists provided $\partial_{\xi_a} B_{bc}-B_{ba}B_{ac}=0$
for each triple of different indices $a,b,c$, but there is no critical point condition when any of the indices 
coincide.

Having a restricted set of fields is part of the reason, mentioned earlier, that our model is not topologically invariant. Indeed, the fields of the form \eqref{eq:B} are not invariant under changes of coordinates, and neither is the critical point condition. Instead there is an infinite dimensional group of 
discrete symmetries arising from the B\"acklund-Moutard type transformations governing the geometry of the generalised 
Darboux system, cf. \cite{Nij2023} and references therein.

The MDC property of the Darboux-KP system, i.e. the relevant integrability aspect, allows us to create a generating 
action for the entire series of higher 
order CS theories, which amounts to a kind of `graded' Chern-Simons  theory in the infinite dimensional space 
of Miwa variables. Thus, 
for the field $\boldsymbol{B}$ of  form \eqref{eq:B}, we can compute at each dimension $2n+1$ the action,
$$\int_{\mathcal{V}_{2n+1}} CS_{2n+1}(\boldsymbol{B}),$$  for each  $2n+1$-dimensional hypersurface $\mathcal{V}_{2n+1}$ embedded in 
$\mathbb{R}^{\mathbb{Z}}$, and consider the formal sum, in powers of a dummy parameter $\hbar$,  
\begin{align*}\label{eq:genCSaction} 
\mathcal{S}_\hbar^{(\infty)}[ \boldsymbol{B}; \mathcal{V}_\infty]= 
\sum_{n=1}^\infty \frac{\hbar^n}{n+1}\int_{\mathcal{V}_{2n+1}}\mathsf{L}^{(2n+1)}.
\end{align*} 
which we could call the \textit{generating Chern-Simons multiform} integrated over the disjoint union 
$\mathcal{V}_\infty=\amalg_{n=1}^\infty \mathcal{V}_{2n+1}$ of submanifolds. 
This novel object makes sense in view of the MDC property of the Darboux-KP system. 
Although it is not clear yet what the physical (or indeed topological) meaning of this infinite-dimensional 
CS theory is, we expect it to play a role in the quantization of the KP system, which,  from different 
perspective, can be studied from the vantage point of the representation theory of 
toroidal algebras along the line of e.g. \cite{Ikeda}.

\section*{Acknowledgement} 
JFM was supported by the EPSRC Programme grant EP/W007509/1: ‘Combinatorial Representation
Theory: Discovering the Interfaces of Algebra with Geometry and Topology’.
FWN was supported by the EPSRC grant EP/W007290/1, when the work was commenced.
DR was supported by the Akroyd \& Brown Scholarship and the School of Mathematics, University of Leeds. JFM would like to thank Ben Lambert for useful discussions.

\noindent \textbf{Data Statement}: No data was created while producing this publication.

\end{document}